\documentclass[12pt]{iopart}
\usepackage{graphicx}
\usepackage{amssymb}

\begin{document}


\title[An extended Landau-Lifshitz-Gilbert equation]{The quantum-mechanical basis of an extended Landau-Lifshitz-Gilbert equation for a current-carrying ferromagnetic wire}


\author{D.M. Edwards$^1$ and O. Wessely$^{1,2}$}
\address{1 Department of Mathematics, Imperial College, London SW7 2BZ, United
 Kingdom}
\address{2 Department of Mathematics, City University,London EC1V 0HB, United 
Kingdom}
\ead{d.edwards@imperial.ac.uk}



\date{\today}

\begin{abstract}
An extended Landau-Lifshitz-Gilbert (LLG) equation is introduced to describe the dynamics of inhomogeneous magnetization in a current-carrying wire. The coefficients of all the terms in this equation are calculated quantum-mechanically for a simple model which includes impurity scattering. This is done by comparing the energies and lifetimes of a spin wave calculated from the LLG equation and from the explicit model. Two terms are of particular importance since they describe non-adiabatic spin-transfer torque and damping processes which do not rely on spin-orbit coupling. It is shown that these terms may have a significant influence on the velocity of a current-driven domain wall and they become dominant in the case of a narrow wall.
\end{abstract}


\maketitle
\section{Introduction}
The effect of passing an electric current down a ferromagnetic wire is of great current interest. If the magnetization is inhomogeneous it experiences a spin-transfer torque due to the current 
\cite{ref:1,ref:2,ref:3,ref:4}. The effect is described phenomenologically by adding terms to the standard LLG equation \cite{ref:5,ref:6}. The leading term in the spin-transfer torque is an adiabatic one arising from that component of the spin polarization of the current which is in the direction of the local magnetization. However , in considering the current-induced motion of a domain wall, Li and Zhang \cite{ref:3,ref:4} found that below a very large critical current the adiabatic term only deforms the wall and does not lead to continuous motion. To achieve this effect they introduced \cite{ref:7} a phenomenological non-adiabatic term associated with the same spin non-conserving processes responsible for Gilbert damping. Subsequently Kohno {\it et al} \cite{ref:8} derived a torque of the Zhang-Li form quantum-mechanically using a model of spin-dependent scattering from impurities. This may arise from spin-orbit coupling on the impurities. More recently Wessely {\it et al} \cite{ref:9} introduced two further non-adiabatic terms in the LLG equation in order to describe their numerical calculations of spin-transfer torques in a domain wall. These quantum-mechanical calculations using the Keldysh formalism were  made  in the ballistic limit without impurities and with spin conserved. Other terms in the LLG equation, involving mixed space and time derivatives, have been considered by Sobolev {\it et al} \cite{ref:11a}, Tserkovnyak {\it et al} \cite{ref:10}, Skadsen {\it et al} \cite{ref:11} and Thorwart and Egger \cite{ref:12}.

The object of this paper is to give a unified treatment of all these terms in the LLG equation and to obtain explicit expressions for their coefficients by quantum-mechanical calculations for a simple one-band  model with and without impurity scattering. The strategy adopted is to consider a uniformly magnetized wire and to calculate the effect of 
a current on the energy and lifetime of a long wavelength spin wave propagating along the wire. It is shown in section 2 that coefficients of spin-transfer torque terms in the LLG equation are directly related to $q$ and $q^3$  terms in the energy and inverse lifetime of a spin wave of wave-vector $q$. The Gilbert damping parameter is the coefficient of the $\omega$  term in the inverse lifetime, where $\omega$ is the spin-wave frequency. It corresponds to the damping of a $q=0$ spin wave while higher order terms $\omega q$ and $\omega q^2$ relate to damping of spin waves with finite wave-vector $q$. The relation between the $q$ term in the spin wave energy and the adiabatic spin-transfer torque has been noticed previously \cite{ref:2,ref:13}. We find that the $q$ term in the spin wave lifetime relates to the Zhang-Li non-adiabatic spin transfer torque. Our result for the coefficient of the Zhang-Li term is essentially the same as that obtained by Kohno {\it et al} \cite{ref:8} and Duine {\it et al} \cite{ref:14} but our derivation appears simpler. The $q^3$ terms in the spin wave energy and lifetime are related to the additional non-adiabatic torques we introduced into the LLG equation \cite{ref:9}, together with an extra one arising from spin non-conserving scattering. Explicit expressions for the coefficients of these terms are obtained in section 3. In section 4 we discuss briefly the importance of the additional terms in our extended LLG equation for current-driven motion of a domain wall. Some conclusions are summarized in section 5.
\section{The LLG equation and spin waves}
We write our extended LLG equation in the dimensionless form
\begin{eqnarray}
{\partial {{\bf s}}\over \partial t}&+&
\alpha{{\bf s}}\times{\partial{{\bf s}}\over \partial t}+
\alpha_1{{\bf s}}\times{\partial^2{{\bf s}}\over \partial z \partial t}- 
\alpha'_1{{\bf s}}\times\left({{\bf s}}\times{\partial^2{{\bf s}}\over \partial z \partial t}\right)
\nonumber \\
&-&\alpha'_2{{\bf s}}\times{\partial^3{{\bf s}}\over \partial z^2 \partial t}- 
\alpha_2{{\bf s}}\times\left({{\bf s}}\times
{\partial^3{{\bf s}}\over\partial z^2 \partial t}\right) 
\nonumber \\
&=&{{\bf s}}\times{\partial^2{{\bf s}} \over \partial z^2}
-b_{ext}{{\bf s}}\times{{\bf e}}_z-a{\partial{{\bf s}}\over \partial z}-f\,{{\bf s}}\times{\partial{{\bf s}}\over \partial z}
\nonumber \\
&+&a_1\left\{{{\bf s}}\times\left({{\bf s}}\times{\partial^3{{\bf s}}\over \partial z^3}\right)+\left[{{\bf s}}\cdot {\partial^2{{\bf s}}\over \partial z^2}-{1\over 2}\left({\partial{{\bf s}}\over \partial z}\right)^2\right]
{\partial{{\bf s}}\over \partial z}\right\}
\nonumber \\
&-&f_1{{\bf s}}\times\left[{{\bf s}}\times{\partial \over \partial z}\left({{\bf s}}\times{\partial^2{{\bf s}}\over \partial z^2}\right)\right]
+g_1{{\bf s}}\times{\partial^3{{\bf s}} \over \partial z^3}.
\label{eq:1}
\end{eqnarray}
Here ${ {\bf s}}(z,t)$ is a unit vector in the direction of the local spin polarisation, time $t$ is measured 
in units of $(\gamma \mu_0 m_s)^{-1}$ and the coordinate $z$ along the wire is in units of the exchange length
$l_{ex}=(2A/\mu_0m_s^2)^{1/2}$. The quantities appearing here are the gyroscopic ratio $\gamma=2\mu_B/\hbar$,the permeability of free space $\mu_0$ and two properties of the ferromagnetic material, namely the saturation
magnetisation $m_s$ and the exchange stiffness constant $A$. ${{\bf e}}_z$ is a unit vector in the 
$z$ direction along the wire. The equation expresses the rate of change of spin angular momentum as the sum of 
various torque terms, of which the $\alpha_1$, $\alpha'_1$, $a$, $f$, $a_1$, $f_1$ and $g_1$ terms are proportional
to the electric current flowing. The second term in the equation is the standard Gilbert term, with damping
factor $\alpha$, while the $\alpha'_1$ and $\alpha'_2$ terms introduce corrections for spin fluctuations of
finite wave-vector. Skadsem {\it et al} \cite{ref:11} point out the existence of the $\alpha'_2$ term but do not
consider it further. It was earlier introduced by Sobolev {\it et al} \cite{ref:11a} within a microscopic
context based on the Heisenberg model. The $\alpha_1$ and $\alpha_2$ terms are found to renormalise the spin wave frequency, but for the model considered in section 3 we find  that $\alpha_1$ is identically zero. We shall argue
that this result is model-independent.  Tserkovnyak {\it et al} \cite{ref:10} and Thorwart and Egger \cite{ref:12} find
non-zero values of $\alpha_1$ which differ from each other by a factor 2; they attribute this to their use of
 Stoner-like and $s-d$ models, respectively. Thorwart and Egger \cite{ref:12} also find the $\alpha'_1$ term and 
they investigate the effect of $\alpha_1$ and $\alpha'_1$ terms on domain wall motion. Their results are 
difficult to assess because the constant $|{{\bf s}}|=1$ is not maintained during the motion. 
In eq.(\ref{eq:1}) we have omitted terms involving the second order time derivatives, whose existence 
was pointed out by Thorwart and Egger \cite{ref:12}; one of these is discussed briefly in section 3.2.

The first term on the right-hand side of eq. (\ref{eq:1}) is due to exchange stiffness and the next term arises from an external magnetic field $B_{ext}{{\bf e}}_z$ with dimensionless coefficient 
$b_{ext}=B_{ext}/\mu_0 m_s$. The third term is the adiabatic spin transfer torque whose coefficient $a$ is simple
 and well-known. In fact \cite{ref:3,ref:4}    
\begin{equation}
 a={1 \over 2} {\hbar JP \over e\mu_0 m^2_sl_{ex}}
\label{eq:2}
\end{equation}
where $J$ is the charge current density and $e$ is the electron charge (a negative quantity). The spin polarisation
factor $P=(J_{\uparrow}-J_{\downarrow})/(J_{\uparrow}+J_{\downarrow})$, where $J_{\uparrow},\;J_{\downarrow}$ are the 
current densities for majority and minority spin in the ferromagnet $(J=J_{\uparrow}+J_{\downarrow})$. 
Eq.(\ref{eq:2}) is valid for both ballistic and diffusive conduction . The fourth term on the right-hand
side of eq.(\ref{eq:1}) is the Zhang-Li torque which is often characterised \cite{ref:8} by a parameter
$\beta=f/a$. The next term is the $E_1$ term of eq.(7) in ref. \cite{ref:9}. It is a non-adiabatic
torque which is coplanar with ${{\bf s}}(z)$ if ${{\bf s}}(z)$ lies everywhere in a plane. As shown in
ref. \cite{ref:9} it is the $z$ derivative of a spin current , which is characteristic of a torque occurring from
spin-conserving processes. In fact this term takes the form 
\begin{equation}
 a_1{\partial \over \partial z}\left[{{\bf s}}\times \left({{\bf s}}\times{\partial^2{{\bf s}}\over \partial z^2}\right)-{1 \over 2} {{\bf s}}\left({\partial{{\bf s}}\over \partial z}\right)^2\right].
\label{eq:2a}
\end{equation}
The $f_1$ term may be written in the form
\begin{equation}
 -f_1\left({{\bf s}}\cdot {\partial {{\bf s}} \over \partial z}\times{\partial^2{{\bf s}}\over \partial z^2}\right){{\bf s}}+
f_1{\partial \over \partial z}\left({{\bf s}}\times{\partial^2{{\bf s}}\over \partial z^2}\right).
\label{eq:3}
\end{equation}
If ${{\bf s}}(z)$ lies in a plane, the case considered in ref. \cite{ref:9}, the first term vanishes and we
recover the $F_1$ term of eq.(9) in ref. \cite{ref:9}. Its derivative form indicate that it arises from spin-conserving processes so we conclude that the coefficient $f_1$ is of that origin. This is not true 
of the last term in eq.(\ref{eq:1}) and we associate the coefficient $g_1$ with spin non-conserving processes. 
For a spin wave solution of the LLG equation, where we work only to first order in deviations from a state of uniform 
magnetisation, the last three terms of eq.(\ref{eq:1}) may be replaced by the simpler ones    
\begin{equation}
 -a_1{\partial^3 {{\bf s}} \over \partial z^3}+\left(f_1+g_1\right){{\bf s}}\times{\partial^3{{\bf s}}\over \partial z^3}.
\label{eq:4}
\end{equation}
Apart from additional terms, eq.(\ref{eq:1}) looks slightly different from eq.(7) of ref. \cite{ref:9} because
we use the spin polarisation unit vector ${{\bf s}}$ rather than the magnetisation vector ${{\bf m}}$
 and ${{\bf s}}=-{{\bf m}}$. Furthermore the dimensionless coefficients will take different numerical values 
because we have used different dimensionless variables $z$ and $t$ to avoid introducing the domain wall width
which was specific to ref. \cite{ref:9}. The torques due to anisotropy fields were also specific
to the domain wall problem and have been omitted in eq.(\ref{eq:1}). 

We suppose that the wire is magnetised
uniformly in the $z$ direction and consider a spin wave as a small transverse oscillation of the spin polarisation 
about the equilibrium state or, when a current flows, the steady state. Thus we look for a solution 
of eq.(\ref{eq:1}) of the form
\begin{equation}
{{\bf s}}=\left(ce^{i(qz-\omega t)},de^{i(qz-\omega t)},-1\right)
\label{eq:5}
\end{equation}
where the coefficients of the $x$ and $y$ components satisfy  $c\ll 1$, $d\ll 1$. This represents a spin wave 
of wave-vector $q$ and angular frequency $\omega$ propagating along the $z$ axis. When (\ref{eq:5}) is substituted
into eq.(\ref{eq:1}) the transverse components yield, to first order in $c$ and $d$, the equations
\begin{equation}
 -i\lambda c +\mu d=0 \,\,\,\,\, , \,\,\,\,\, \mu c+ i\lambda d=0
\label{eq:6}
\end{equation}
where
\begin{eqnarray}
\lambda &=& \omega-aq+a_1q^3 -\alpha_2\omega q^2 +i\alpha'_1q\omega
\nonumber \\
\mu &=& -i\alpha \omega + b_{ext} +q^2+ifq+i(f_1+g_1)q^3
+\alpha_1\omega q-i\omega q^2 \alpha'_2.
\label{eq:7}
\end{eqnarray}
On eliminating $c$ and $d$ from eq.(\ref{eq:6}) we obtain  $\lambda^2=\mu^2$. To obtain a positive real part 
for the spin wave frequency, we take $\lambda=\mu$. Hence 
\begin{eqnarray}
&&\omega \left(1-\alpha_1 q-\alpha_2 q^2\right)
\nonumber \\
&&=b_{ext}+aq+q^2-a_1q^3
\nonumber \\
&&+i\left[\omega\left(-\alpha-\alpha'_1q-\alpha'_2q^2\right)
+fq+\left(f_1+g_1\right)q^3\right].
\label{eq:8}
\end{eqnarray}
Thus the spin wave frequency is given by
\begin{equation}
 \omega=\omega_1-i\omega_2
\label{eq:9}
\end{equation}
where
\begin{eqnarray}
 \omega_1&\simeq&\left(1-\alpha_1 q-\alpha_2 q^2\right)^{-1}
\left(b_{ext}+aq+q^2-a_1q^3\right)
\nonumber \\
\omega_2&\simeq&\left(1-\alpha_1 q-\alpha_2 q^2\right)^{-1}\left[\omega_1\left(\alpha+\alpha'_1q+\alpha'_2q^2\right)
-fq-\left(f_1+g_1\right)q^3\right].
\label{eq:10}
\end{eqnarray}
Here we have neglected terms of second order in $\alpha$,
$\alpha'_1$, $\alpha'_2$, $f$, $f_1$ and $g_1$, the coefficients which appear in the spin wave damping.
This form for the real and imaginary parts of the spin wave frequency is convenient for comparing with the
quantum-mechanical results of the next section. In this way we shall obtain explicit expressions for
all the coefficients in the phenomenological LLG equation. Coefficients of odd powers of $q$ are proportional
to the current flowing whereas terms in even powers of $q$ are present in the equilibrium state with zero current. 
\section{Spin wave energy and lifetimes in a simple model}
As a simple model of an itinerant electron ferromagnet we consider the one-band Hubbard model
\begin{equation}
 H_0=-t\sum_{ij\sigma}c_{i\sigma}^{\dag}c_{j\sigma}+U\sum_i n_{i\uparrow}n_{i\downarrow}
-\mu_B B_{ext}\sum_{i}\left(n_{i\uparrow}-n_{i\downarrow}\right),
\label{eq:11}
\end{equation}
where $c_{i\sigma}^{\dag}$ creates an electron on site $i$ with spin $\sigma$ and $n_{i\sigma}=c_{i\sigma}^{\dag}c_{i\sigma}$. We consider a simple cubic lattice and the intersite
 hopping described by the first term is restricted to nearest neighbours. The second term describes
 an on-site interaction between electrons with effective interaction parameter $U$; the last term is
 due to an external magnetic field. It is convenient to introduce a Bloch representation, with  
\begin{eqnarray}
 c_{{\bf k}\sigma}^\dag={1 \over \sqrt{N}}\sum_ie^{{\bf k}\cdot {\bf R}_i} c_{i\sigma}^{\dag} \,\,\,\,\, , \,\,\,\,\, n_{{\bf k}\sigma}=c_{{\bf k}\sigma}^{\dag}c_{{\bf k}\sigma},
\label{eq:12a} \\
\epsilon_{\bf k}=-t\sum_ie^{i{\bf k}\cdot {\bf \rho}_i}=-2t\left(\cos k_x a_0+\cos k_y a_0+\cos k_z a_0 \right).
\label{eq:12b}
\end{eqnarray} 
The sum in eq.(\ref{eq:12a}) is over all lattice cites ${\bf R}_i$ whereas in eq.(\ref{eq:12b}) 
${\bf \rho}_i=(\pm a_0,0,0),(0,\pm a_0,0),(0,0,\pm a_0)$ 
are the nearest neighbour lattice sites. Then 
\begin{equation}
 H_0=\sum_{{\bf k} \sigma}\epsilon_{\bf k} n_{{\bf k}\sigma}+U\sum_{i}n_{i\uparrow}n_{i\downarrow}-\mu_B B_{ext}\sum_{\bf k}\left(n_{{\bf k}\uparrow}-n_{{\bf k}\downarrow}\right).
\label{eq:13}
\end{equation}
To discuss scattering of spin waves by dilute impurities we assume that the effect of the scattering
 from different impurity sites adds incoherently; hence we may consider initially a single scattering center
 at the origin, We therefore introduce at this site a perturbing potential $u+v{\bf l} \cdot {\bf \sigma}$,
 where
${\bf l}=(\sin \theta \cos \phi, \sin \phi \sin \theta, \cos \theta)$ is a unit vector whose direction
 will finally be averaged over. $u$ is the part of the impurity potential which is indepndent of the
 spin $ {\bf \sigma}$ and the  spin dependent potential $v{\bf l}\cdot {\bf \sigma}$ is intended to
 simulate a spin-orbit ${\bf L}\cdot{\bf \sigma}$ interaction on the impurity. It breaks spin rotational
 symmetry in the simplest possible way. Clearly spin-orbit coupling can only be treated correctly for
 a degenerate band such as a $d$-band, where on-site orbital angular momentum ${\bf L}$ occurs naturally.
 The present model is equivalent to that used by Kohno {\it et al} \cite{ref:8} and Duine {\it et al} \cite{ref:14}. In Bloch representation
 the impurity potential becomes $V=V_1+V_2$ with
\begin{eqnarray}
 V_1&=&v_{\uparrow}{1\over N}\sum_{{\bf k}_1{\bf k}_2}c_{{\bf k}_1\uparrow}^\dag c_{{\bf k}_2\uparrow}
+v_{\downarrow}{1\over N}\sum_{{\bf k}_1{\bf k}_2}c_{{\bf k}_1\downarrow}^\dag c_{{\bf k}_2\downarrow}
\nonumber \\
 V_2&=&ve^{-i\phi}\sin \theta {1\over N}\sum_{{\bf k}_1{\bf k}_2}c_{{\bf k}_1\uparrow}^\dag c_{{\bf k}_2\downarrow}
+ve^{i\phi}\sin \theta {1\over N}\sum_{{\bf k}_1{\bf k}_2}c_{{\bf k}_1\downarrow}^\dag c_{{\bf k}_2\uparrow}
\label{eq:15}
\end{eqnarray}
and $v_{\uparrow}=u+v\cos \theta$ , $v_{\downarrow}=u-v\cos \theta$.
To avoid confusion we note that the spin dependence of the impurity potential which occurs in the
 many-body Hamiltonian $H_0+V$ is not due to exchange, as would arise in an approximate self consistent
 field treatment (e.g. Hartree-Fock) of the interaction  $U$ in a ferromagnet.

\subsection{Spin wave energy and wave function}

In this section we neglect the perturbation due to impurities and determine expressions for the
 energy and wave function of a long-wave length spin wave in the presence of an electric current.
 The presence of impurities is recognised implicitly since the electric current is characterised
 by a perturbed one-electron distribution function $f_{{\bf k}\sigma}$ which might be obtained by solving a Boltzman equation with a collision term. We consider a spin wave of wave-vector ${\bf q}$ propagating along
the $z$ axis, which is the direction of current flow. Lengths and times used in this section and the next, except 
when specified, correspond to actual physical quantities, unlike  the dimensionless variables used 
in section 2.  

We first consider the spin wave with zero electric current and treat it, within the random phase approximation
(RPA), as an excitation from the Hartree-Fock (HF) ground state of the Hamiltonian (\ref{eq:13}). The HF one electron energies are given by
\begin{equation}
E_{{\bf k}\sigma}=\epsilon_{\bf k}+U\left<n_{-\sigma}\right>-\mu_B\sigma B_{ext}
\label{eq:16}
\end{equation}
where $\sigma=1,-1$ for $\uparrow$ and $\downarrow$ respectively, and  $\langle n_{-\sigma}\rangle$ is the 
number of $-\sigma$ spin electrons per site. In a self-consistent ferromagnetic state at $T=0$, 
$\langle n_{\sigma}\rangle=N^{-1}\sum_{\bf k}f_{{\bf k}\sigma}$ and $n=\sum_{\sigma}\langle n_{\sigma}\rangle$,
where, $f_{{\bf k}\sigma}=\theta(E_F-E_{{\bf k}\sigma})$, $n$ is the number of electrons per atom, and $E_F$ is the Fermi energy. $N$ is the number of lattice sites and $\theta(E)$ is the unit step function.
The spin bands $E_{{\bf k}\sigma}$ given by eq.(\ref{eq:16}) are shifted relative to each other by an energy 
$\Delta+2\mu_B B_{ext}$ where $\Delta=U\langle n_{\uparrow}-n_{\downarrow}\rangle$ is the exchange splitting.
The ground state is given by $|0\rangle=\prod_{{\bf k}\sigma}c^{\dag}_{{\bf k}\sigma}|\rangle$ where $|\rangle$
is the vacuum state and the product extends over all states ${\bf k}\sigma$ such that $f_{{\bf k}\sigma}=1$.
Within the RPA, the wave function for a spin wave of wave-vector ${\bf q}$, excited from the HF ground state,
takes the form
\begin{equation}
 \left|{\bf q} \right>=N_{\bf q}\sum_{\bf k}A_{\bf k}c_{{\bf k}+{\bf q}\downarrow}^\dag
c_{{\bf k}\uparrow}\left|0\right>
\label{eq:17}
\end{equation}
where $N_{\bf q}$ is a normalisation factor. The energy of this state may be written
\begin{equation}
 E_{\bf q}=E_{gr}+\hbar\omega_{\bf q}=E_{gr}+2\mu_B B_{ext}+\hbar\omega'_{\bf q}
\label{eq:18}
\end{equation}
where $E_{gr}$ is the energy of the HF ground state and $\hbar \omega_{\bf q}$ is the spin wave
excitation energy.
On substituting (\ref{eq:17}) in the Schr\"{o}dinger equation $(H_0-E_{\bf q})|{\bf q}\rangle=0$ and
multiplying on the left by $\langle 0|c_{{\bf k}'\uparrow}^{\dagger}c_{{\bf k}'-{\bf q}\downarrow}$, we find
\begin{equation}
 A_{{\bf k}'}\left(\epsilon_{{\bf k}'+{\bf q}}-\epsilon_{{\bf k}'}+\Delta -\hbar \omega'_{\bf q}\right)=
{U \over N}\sum_{\bf k}A_{\bf k}f_{{\bf k}\uparrow}\left(1-f_{{\bf k}+{\bf q}\downarrow}\right).
\label{eq:19}
\end{equation}
Hence we may take  
\begin{equation}
 A_{{\bf k}}=\Delta\left(\epsilon_{{\bf k}+{\bf q}}-\epsilon_{\bf k}+\Delta -\hbar \omega'_{\bf q}\right)^{-1}
\label{eq:20}
\end{equation}
and, for small ${\bf q}$, $\hbar \omega'_{\bf q}$ satisfies the equation 
\begin{equation}
 1={U \over N}\sum_{\bf k}{f_{{\bf k}\uparrow}-f_{{\bf k}+{\bf q} \downarrow} \over \epsilon_{{\bf k}+{\bf q}}-\epsilon_{\bf k}+\Delta -\hbar \omega'_{\bf q}}.
\label{eq:21}
\end{equation}
This is the equation for the poles of the well-known RPA dynamical susceptibility $\chi({\bf q},\omega)$
\cite{ref:14a}. The spin wave pole is the one for which $\hbar \omega'_{\bf q}\rightarrow 0$ as 
${\bf q}\rightarrow 0$. 

To generalise the above considerations to a current-carrying state we proceed 
as follows. We re-interpret the state $|0\rangle$ such that $\langle 0|\ldots|0\rangle$ corresponds 
to a suitable ensemble average with a modified one-electron distribution $f_{{\bf k}\sigma}$. When a current
flows in the $z$ direction we may consider the $\uparrow$ and $\downarrow$ spin Fermi surfaces as shifted
by  small displacement $\delta_{\uparrow}{\hat{\bf k}}_z$, $\delta_{\downarrow}{\hat{\bf k}}_z$ where
${\hat{\bf k}}_z$ is a unit vector in the $z$ direction. Thus
\begin{eqnarray}
f_{{\bf k}\sigma}&=&\theta(E_F-E_{{\bf k}+\delta_{\sigma}{\hat{\bf k}}_z,\sigma})
\nonumber \\
&\simeq& \theta(E_F-E_{{\bf k}\sigma})-\delta_{\sigma}\delta(E_F-E_{{\bf k}\sigma}){\partial \epsilon_{\bf k} \over \partial k_z}
\label{eq:22}
\end{eqnarray}
and the charge current density carried by spin $\sigma$ electrons is
\begin{eqnarray}
 J_{\sigma}&=&{e \over \hbar N a_0^3}\sum_{\bf k}{\partial \epsilon_{\bf k} \over \partial k_z}f_{{\bf k}\sigma}=
-{e \delta_{\sigma}\over \hbar N a_0^3}\sum_{\bf k}\left({\partial \epsilon_{\bf k} \over \partial k_z}\right)^2\delta(E_F-E_{{\bf k}\sigma})
\nonumber \\
&=&-{e \delta_{\sigma}\over \hbar a_0^3}\left<\left({\partial \epsilon_{\bf k} \over \partial k_z}\right)^2\right>_{\sigma}\rho_{\sigma}(E_F)
\label{eq:23}
\end{eqnarray}
where $\langle(\partial\epsilon_{\bf k}/\partial k_z)^2\rangle_{\sigma}$ is an average over the 
$\sigma$ spin Fermi surface and $\rho_{\sigma}(E_F)$ is the density of $\sigma$ spin states per atom at the
Fermi energy. We shall also encounter the following related quantities; 
\begin{eqnarray}
 K_{\sigma}&=&{ 1 \over N \Delta^2 a_0^3}\sum_{\bf k}{\partial \epsilon_{\bf k} \over \partial k_z}{\partial^2 \epsilon_{\bf k} \over \partial k^2_z}f_{{\bf k}\sigma}
\nonumber \\
&=&{\hbar J_{\sigma} \over \Delta^2 e}\left<\left({\partial \epsilon_{\bf k} \over \partial k_z}\right)^2{\partial^2 \epsilon_{\bf k} \over \partial k^2_z}\right>_{\sigma}\Bigg/\left<\left({\partial \epsilon_{\bf k} \over \partial k_z}\right)^2\right>_{\sigma}
\label{eq:24}\\
L_{\sigma}&=&{ 1 \over N \Delta^3 a_0^3}\sum_{\bf k}\left({\partial \epsilon_{\bf k} \over \partial k_z}\right)^3f_{{\bf k}\sigma}
\nonumber \\
&=&{\hbar J_{\sigma} \over \Delta^3 e}\left<\left({\partial \epsilon_{\bf k} \over \partial k_z}\right)^4\right>_{\sigma}\Bigg/\left<\left({\partial \epsilon_{\bf k} \over \partial k_z}\right)^2\right>_{\sigma}.
\label{eq:25}
\end{eqnarray}
To derive eqs.(\ref{eq:24}) and (\ref{eq:25}), $\delta_{\sigma}$ has been eliminated using eq.(\ref{eq:23}).

To solve eqn.(\ref{eq:21}) for $\hbar \omega_{\bf q}'$ we expand the right-hand side of the equation in powers
of $(\epsilon_{{\bf k}+{\bf q}}-\epsilon_{\bf k}-\hbar \omega_{\bf q}')/\Delta$ and make the further expansions
\begin{eqnarray}
 \epsilon_{{\bf k}+{\bf q}}-\epsilon_{\bf k}&=&q{\partial \epsilon_{\bf k} \over \partial k_z}+{1 \over 2}q^2{\partial^2 \epsilon_{\bf k} \over \partial k^2_z}
+{1 \over 6}q^3{\partial^3 \epsilon_{\bf k} \over \partial k^3_z} \ldots
\label{eq:26}\\
\hbar \omega'_{\bf q}&=&Bq+Dq^2+Eq^3+\ldots
\label{eq:27}
\end{eqnarray}
in powers of $q$. We retain all terms up to $q^3$ except those involving $B^{2}$; the coefficients  $B$ and $E$
are proportional to the current and we keep only terms linear in the current. Hence we find a solution 
of eq.(\ref{eq:21}) in the form (\ref{eq:27}) with 
\begin{eqnarray}
 B&=&{ 1 \over N_{\uparrow}-N_{\downarrow}}\sum_{\bf k}\left(f_{{\bf k}\uparrow}-f_{{\bf k}\downarrow}\right){\partial \epsilon_{\bf k} \over \partial k_z}=
{Na^3_0 \over  N_{\uparrow}-N_{\downarrow}}{\hbar \over e}\left(J_{\uparrow}-J_{\downarrow}\right)
\label{eq:28}\\
D&=&{ 1 \over N_{\uparrow}-N_{\downarrow}}\left[{1 \over 2}\sum_{\bf k}\left(f_{{\bf k}\uparrow}+f_{{\bf k}\downarrow}\right){\partial^2 \epsilon_{\bf k} \over \partial k^2_z}-{1 \over \Delta}\sum_{\bf k}\left(f_{{\bf k}\uparrow}-f_{{\bf k}\downarrow}\right)\left({\partial \epsilon_{\bf k} \over \partial k_z}\right)^2\right]
\label{eq:29} \\
E&=&-{a^2_0B \over 6}
\nonumber \\
&+&{ B \over \left(N_{\uparrow}-N_{\downarrow}\right)\Delta}
\left[\sum_{\bf k}\left(f_{{\bf k}\uparrow}+f_{{\bf k}\downarrow}\right){\partial^2 \epsilon_{\bf k} \over \partial k^2_z}-{3 \over \Delta}\sum_{\bf k}\left(f_{{\bf k}\uparrow}-f_{{\bf k}\downarrow}\right)\left({\partial \epsilon_{\bf k} \over \partial k_z}\right)^2\right]
\nonumber \\
&-&U a^3_0\sum_{\sigma}\left(K_{\sigma}-\sigma L_{\sigma}\right).
\label{eq:30}
\end{eqnarray}
Here $N_{\sigma}$ is the total number of $\sigma$ spin electrons so that 
$N_{\sigma}=N\langle n_{\sigma}\rangle$. 

In the absence of spin-orbit coupling  the expression for $B$ in terms of spin current
is a general exact result even in the presence of disorder, as shown in Appendix A.
The coefficient $D$ is the standard RPA spin-wave stiffness constant
(e.g ref. \cite{ref:14a}). We note that, in the limit $\Delta\rightarrow\infty$, E takes the simple form 
$-a_0^2B/6$.

On restoring the correct dimensions (as indicated after eq.(\ref{eq:1})) to the expression for $\omega_1$
in eq.(\ref{eq:10}) we may determine the coefficients $a$ and $a_1$ by comparing with the equation   
\begin{equation}
 \hbar \omega_{\bf q}=2\mu_B B_{ext}+Bq+Dq^2+Eq^3.
\label{eq:31}
\end{equation}
From the coefficient of $q$ we have
\begin{equation}
 a+\alpha_1b_{ext}=B/\left(2\mu_B \mu_0 m_s l_{ex}\right).
\label{eq:32}
\end{equation}
$a$ and $B$ are both determined directly from the spin current $JP$ independently of a particular model
(see appendix A) so that  $b_{ext}$ should not enter their relationship. We conclude quite generally that
$\alpha_1=0$. In this case we find that on combining eqs.(\ref{eq:32}) and (\ref{eq:28}), and noting
that $m_s=-\mu_B(N_{\uparrow}-N_{\downarrow})/Na_0^3$, eq.(\ref{eq:2}) is obtained as expected. In section
3.2 we show explicitly for the present model that $\alpha_1=0$. This conflicts with the results of 
refs.\cite{ref:10} and \cite{ref:12}. From the coefficients of $q^2$ in eqs.(\ref{eq:10}) and
(\ref{eq:31}) we find $1+\alpha_2b_{ext}=D/(4\mu_BA/m_s)$. Thus an external field slightly disturbs
the standard relation $A=Dm_s/4\mu_B$. However in the spirit of the LLG equation we take $A$ and $m_s$, 
which enter the units of length and time used in eq.(\ref{eq:1}), to be constants of the ferromagnetic
material in zero external field. The coefficients of $q^3$ in eqs.(\ref{eq:10}) and (\ref{eq:31}) yield the 
relation (taking $\alpha_1=0$),  
\begin{equation}
 -a_1+\alpha_2a=E/\left(2\mu_B \mu_0 m_s l^3_{ex}\right).
\label{eq:33}
\end{equation}
We defer calculation of $\alpha_2$ until section 3.2 and the result is given in eq.(\ref{eq:43}). 
Combining this with eqs.(\ref{eq:33}) and (\ref{eq:30}) we find
\begin{equation}
 2\mu_B \mu_0 m_s l^3_{ex}a_1={a^2_0 B \over 6}- {2BD \over \Delta}+Ua_0^3\sum_{\sigma}\left(K_{\sigma}-\sigma L_{\sigma}\right).
\label{eq:34}
\end{equation}
We have thus derived an explicit expression , for a simple model, for the coefficient $a_1$ of a non-adiabatic
spin torque term which appears in the LLG equation (\ref{eq:1}). We have neglected the effect of disorder
due to impurities . In the absence of spin-orbit coupling the expression for the adiabatic torque coefficient 
$a$, given by eq.(\ref{eq:2}), is exact even in presence of impurities. In the next section we shall calculated further non-adiabatic torque terms, with coefficients $f_1$ and $g_1$, as well as damping coefficients $\alpha$, $\alpha'_1$ and $\alpha'_2$. In the present model all these depend on impurity
scattering for their existence.

\subsection{Spin wave lifetime}
	  \begin{figure}
	    \includegraphics[width=0.6\textwidth]{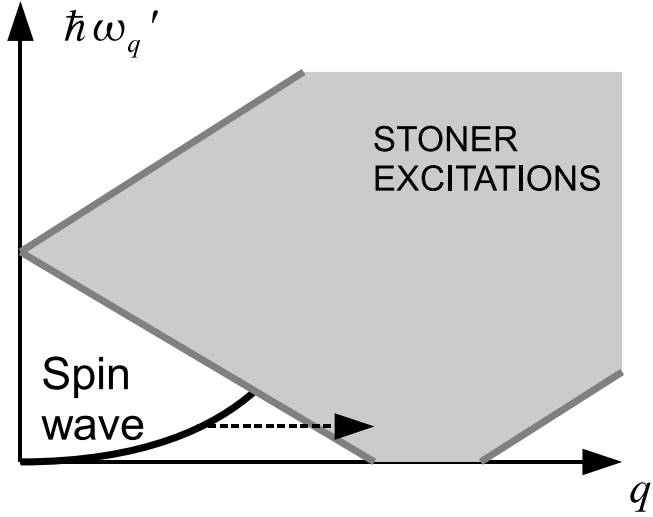}
	    \caption{\label{fig:1} Spin-flip excitations from the ferromagnetic ground state.
			The dotted arrow shows the mechanism of decay of a spin wave into Stoner excitations
			which is enabled by the impurity potential $V_1$.}
	  \end{figure}
The solutions of eq.(\ref{eq:21}) are shown schematically in figure \ref{fig:1}. They include the spin wave dispersion curve and the continuum of Stoner excitations 
$c^{\dag}_{{\bf k}+{\bf q}\downarrow}c_{{\bf k}\uparrow}|0\rangle$ with energies $E_{{\bf k}+{\bf q}\downarrow}-E_{{\bf k}\uparrow}$. The Zeeman gap $2\mu_B B_{ext}$ in the spin wave energy at 
${\bf q}=0$ does not appear because we have plotted $\hbar \omega'_{\bf q}$ rather than  
$\hbar \omega_{\bf q}$ (see eq.(\ref{eq:18})). Within the present RPA the spin wave in a pure
metal has infinite lifetime outside the continuum and cannot decay into Stoner excitations owing to conservation
of the momentum $\bf q$. However, when the perturbation $V_1$ due to impurities is introduced 
(see eqn.(\ref{eq:15})), crystal momentum is no longer conserved and such decay processes can occur. These
are shown schematically by the dotted arrow in figure \ref{fig:1}. If the bottom of the $\downarrow$ spin 
band lies above the Fermi level there is a gap in the Stoner spectrum and for a low energy (small $q$) spin wave such
processes cannot occur. However the spin-flip potential $V_2$ enables the spin wave to decay into single particle excitations $c^{\dag}_{{\bf k}+{\bf q}\sigma}c_{{\bf k}\sigma}|0\rangle$ about each Fermi surface
and these do not have an energy gap. 

The lifetime $\tau^{-1}_{\bf q}$ of a spin wave of wave-vector 
${\bf q}$ is thus given simply by the ``golden rule'' in the form
\begin{equation}
 \tau_q^{-1}={2\pi \over \hbar}N_{imp}\left(T_1+T_2\right)
\label{eq:35}
\end{equation}
where $N_{inp}$ is the number of impurity sites and
\begin{eqnarray}
 T_1&=&\sum_{{\bf k}{\bf p}}\left|\left<0\left|c_{{\bf k}\uparrow}^{\dag}c_{{\bf p}\downarrow}V_1\right|{\bf q}\right>\right|^2
f_{{\bf k}\uparrow}\left(1-f_{{\bf p}\downarrow}\right)\delta\left(\hbar \omega_{\bf q}-E_{{\bf p}\downarrow}+E_{{\bf k}\uparrow}\right)
\nonumber \\
 T_2&=&\sum_{{\bf k}{\bf p}\sigma}\left|\left<0\left|c_{{\bf k}\sigma}^{\dag}c_{{\bf p}\sigma}V_2\right|{\bf q}\right>\right|^2
f_{{\bf k}\sigma}\left(1-f_{{\bf p}\sigma}\right)\delta\left(\hbar \omega_{\bf q}-\epsilon_{{\bf p}}+\epsilon_{\bf k}\right).
\label{eq:36}
\end{eqnarray}
We first consider $T_1$ and, using eqns.(\ref{eq:15}) and (\ref{eq:17}), we find
\begin{eqnarray}
 \left<0\left|c_{{\bf k}\uparrow}^{\dag}c_{{\bf p}\downarrow}V_1\right|{\bf q}\right>
&=&{N_{\bf q} \over N}f_{{\bf k}\uparrow}\left(1-f_{{\bf p}\downarrow}\right)
\left[A_{\bf k}v_{\downarrow}\left(1-f_{{\bf p}\downarrow}\right)-A_{{\bf p}-{\bf q}}v_{\uparrow}f_{{\bf p}-{\bf q}\uparrow} \right]
\nonumber \\
&=&{N_{\bf q} \over N}f_{{\bf k}\uparrow}\left(1-f_{{\bf p}\downarrow}\right)\left(A_{\bf k}v_{\downarrow}-A_{{\bf p}-{\bf q}}v_{\uparrow}\right)
\label{eq:37}
\end{eqnarray}
for small $q$. The last line follows from two considerations. Firstly, because of the $\delta$-function in
eq.(\ref{eq:36}) we can consider the states ${\bf k}_{\uparrow}$ and ${\bf p}_{\downarrow}$ to be close
to their respective Fermi surfaces. Secondly the $\downarrow$ spin Fermi surface lies within the 
$\uparrow$ Fermi surface and $q$ is small. Hence
\begin{equation}
 T_1={N^2_{\bf q} \over N^2}\sum_{{\bf k}{\bf p}}f_{{\bf k}\uparrow}\left(1-f_{{\bf p}\downarrow}\right)\delta\left(\hbar \omega_{\bf q}-E_{{\bf p}\downarrow}+E_{{\bf k}\uparrow}\right)\left(A_{\bf k}v_{\downarrow}-A_{{\bf p}-{\bf q}}v_{\uparrow}\right)^2.
\label{eq:38}
\end{equation}
To evaluate this expression in the case when a current flows we use the distribution function  
$f_{{\bf k}\sigma}$ given by eq.(\ref{eq:22}). Thus, neglecting a term proportional to the square of
the current, we have
\begin{eqnarray}
T_1&=&{N^2_{\bf q} \over N^2}\sum_{{\bf k}{\bf p}}\delta\left(\hbar \omega_{\bf q}-E_{{\bf p}\downarrow}+E_{{\bf k}\uparrow}\right)\left(A_{\bf k}v_{\downarrow}-A_{{\bf p}-{\bf q}}v_{\uparrow}\right)^2
\nonumber \\
&\times&\Bigg[\theta(E_F-E_{{\bf k}\uparrow})\theta(E_{{\bf p}\downarrow}-E_F)
-\delta_{\uparrow}\theta(E_{{\bf p}\downarrow}-E_F)\delta(E_F-E_{{\bf k}\uparrow}){\partial \epsilon_{\bf k} \over \partial k_z}
\nonumber \\
&+&\delta_{\downarrow}\theta(E_F-E_{{\bf k}\uparrow})\delta(E_F-E_{{\bf p}\downarrow}){\partial \epsilon_{\bf p} \over \partial p_z}\Bigg].
\label{eq:39}
 \end{eqnarray}
We wish to expand this expression, and a similar one for $T_2$, in powers of $q$ to $O(q^3)$ so that we can
compare with the phenomenological expression (eq.(\ref{eq:10})) for the imaginary part of the spin wave frequency, which is given by $\tau^{-1}_{\bf q}/2$. It is straight-forward to expand 
the second factor in the above sum by using eqs.(\ref{eq:20})
and (\ref{eq:27}). We shall show that the contribution to $T_1$ of the first term in square
brackets in eq.(\ref{eq:39}) leads to a contribution proportional to spin wave frequency $\omega_{\bf q}$.
Together with a similar contribution to $T_2$ it yields the Gilbert damping factor $\alpha$ as well as the 
coefficients $\alpha'_1$, $\alpha'_2$ of the terms in eq.(\ref{eq:10}) which give the ${\bf q}$ dependence
of the damping. The remaining terms in eq.(\ref{eq:39}) yield the spin-transfer torque coefficients
$f$, $f_1$ and $g_1$.

The normalisation factor $N^2_{\bf q}$ which appear in eq.(\ref{eq:39}) leads naturally to the factor
$(1-\alpha_1q-\alpha_2q^2)^{-1}$ which appears in eq.(\ref{eq:10}). From eq.(\ref{eq:17}) it is given by
\begin{equation}
 1=\left<{\bf q}|{\bf q}\right>={N^2_{\bf q} \over N}
\sum_{\bf k}\left(A^2_{\bf k}f_{{\bf k}\uparrow}-A^2_{{\bf k}-{\bf q}}f_{{\bf k}\downarrow}\right).
\label{eq:40}
\end{equation}
By expanding $A^2_{{\bf k}-{\bf q}}$ in powers of $q$, and using eq.(\ref{eq:22}), we find to $O(q^2)$ that
\begin{eqnarray}
 N_{\bf q}^{-2}=\left(N_{\uparrow}-N_{\downarrow}\right)
\nonumber \\
\times \left\{1+{{\bf q}^2 \over \Delta^2\left(N_{\uparrow}-N_{\downarrow}\right)}
\sum_{\bf k}\left({\partial \epsilon_{\bf k} \over \partial k_z}\right)^2
\left[\theta(E_F-E_{{\bf k}\uparrow})-\theta(E_F-E_{{\bf k}\downarrow})\right]\right\}.
\label{eq:41}
\end{eqnarray}
We deduce that
\begin{equation}
 \alpha_1=0
\label{eq:42}
\end{equation}
and
\begin{eqnarray}
\alpha_2=-{1 \over l^2_{ex}\Delta^2\left(N_{\uparrow}-N_{\downarrow}\right)}
\sum_{\bf k}\left({\partial \epsilon_{\bf k} \over \partial k_z}\right)^2
\left[\theta(E_F-E_{{\bf k}\uparrow})-\theta(E_F-E_{{\bf k}\downarrow})\right].
\label{eq:43}
\end{eqnarray}
The result $\alpha_1=0$, which was predicted on general grounds in section 3.1 and in Appendix 1, arises here through the absence
of a $q$ term, proportional to current, in the spin wave normalisation factor. In the derivation 
of eq.(\ref{eq:41}) this occurs due  to a cancellation involving the $Bq$ terms in the spin energy, which appears in $A_{\bf k}$. Without this cancellation we would have $\alpha_1=2B/l_{ex} \Delta$ which is of the form obtained
by Tserkovnyak {\it et al} \cite{ref:10} and Thorwart and Egger \cite{ref:12}.
 
We now return to the programme for calculating the LLG coefficients $\alpha,\alpha'_1,\alpha'_2,f,f_1,g_1$ 
which was outlined after eq.(\ref{eq:39}). We have seen that the $q$ dependence of $N^2_{q}$ corresponds to
the prefactor in eq.(\ref{eq:10}). Hence to determine the coefficients listed above we can take 
$N^2_{q}=N^2_0=(N_{\uparrow}-N_{\downarrow})^{-1}$ in $T_1$ and $T_2$ when we expand terms in powers of $q$
to substitute in eq.(\ref{eq:35}) and compare with eq.(\ref{eq:10}). We first consider the case $q=0$ in
order to determine the Gilbert damping factor $\alpha$. Thus only the first term in square brackets in 
eq.(\ref{eq:39}) contributes, since $\partial \epsilon_{\bf k} /\partial k_z$ is an odd function $k_z$, and
\begin{eqnarray}
 T_1({\bf q}=0)&=&{4v^2\overline{\cos^2\theta} \over N_{\uparrow}-N_{\downarrow}}
\nonumber \\
&\times& N^{-2}\sum_{{\bf k}{\bf p}}
\delta\left(\hbar \omega_0-E_{{\bf p}\downarrow}+E_{{\bf k}\uparrow}\right)
\theta(E_F-E_{{\bf k}\uparrow})\theta(E_{{\bf p}\downarrow}-E_F)
\label{eq:44}
\end{eqnarray}
where $\overline{\cos^2\theta}$ is an average over the angle appearing in the impurity potential $V$ 
(eq.(\ref{eq:15})) and we shall assume $\overline{\cos\theta}=0$. The summations in eq.(\ref{eq:44}) may be replaced
by energy integrals involving the density of states of per atom $\rho_{\sigma}(\epsilon)$ of the states
$E_{{\bf k}\sigma}$. Then, to order $(\hbar \omega_0)^2$,
\begin{equation}
 T_1({\bf q}=0)=\left[{4v^2\overline{\cos^2\theta}\over N_{\uparrow}-N_{\downarrow}}\right]\left[\hbar \omega_0 \rho_{\uparrow} \rho_{\downarrow}+{1 \over 2}\left(\hbar \omega_0\right)^2
\left(\rho_{\uparrow} \rho'_{\downarrow}-\rho'_{\uparrow} \rho_{\downarrow}\right)\right]
\label{eq:45}
\end{equation}
where $\rho_{\sigma}(\epsilon)$ and its derivative $\rho'_{\sigma}(\epsilon)$ are evaluated at $\epsilon=E_F$.
Similarly
\begin{equation}
 T_2({\bf q}=0)=\left[{v^2\overline{\sin^2\theta}\over N_{\uparrow}-N_{\downarrow}}\right]\hbar \omega_0 \left(\rho^2_{\uparrow} +\rho^2_{\downarrow}\right)
\label{eq:46}
\end{equation}
and no $\omega_0^2$ terms appear. We have included the $\omega_0^2$ term in eq.(\ref{eq:45}) merely because
it corresponds to a term ${{\bf s}}\times\left({{\bf s}}\times{\partial^2{{\bf s}}\over  \partial t^2}\right)$ in the LLG equation whose existence was noted by Thorwald and Egger \cite{ref:12}. We shall not
pursue terms with second-order time derivatives any further. Since the imaginary part of the spin wave frequency 
is given by $\tau^{-1}_{q}/2$ it follows from eqs.(\ref{eq:10}), (\ref{eq:35}), (\ref{eq:45}) and (\ref{eq:46})
that
\begin{equation}
 \alpha={\pi c v^2 \over \left<n_{\uparrow}-n_{\downarrow}\right>}
\left[4\overline{\cos^2\theta}\rho_{\uparrow} \rho_{\downarrow}
+\overline{\sin^2\theta}\left(\rho^2_{\uparrow} +\rho^2_{\downarrow}\right)\right],
\label{eq:47}
\end{equation}
where $c=N_{imp}/N$ is the concentration of impurities, in agreement with Khono {\it et al} \cite{ref:8}
and Duine {\it et al} \cite{ref:14}. If the direction of the spin quantisation axis of the impurities is distributed randomly $\overline{\cos^2\theta}=1/3,\overline{\sin^2\theta}=2/3$ so that $\alpha$ is proportional to
$(\rho_{\uparrow}+\rho_{\downarrow})^2$.

To investigate the ${\bf q}$ dependence of Gilbert damping, and thus evaluate $\alpha'_1$ and $\alpha'_2$ in 
eq.(\ref{eq:10}), the second factor in the summation of eq.(\ref{eq:39}) must be expanded in powers of $q$.
All the terms which contribute to the sum are of separable form $g({\bf k})h({\bf p})$. The contribution to 
$T_1$ of interest here , proportional to $\omega_{\bf q}$, again arises from the first term in square brackets in eq.(\ref{eq:39}), and similarly for $T_2$. The summations required in eq.(\ref{eq:39}) are of the form
\begin{eqnarray}
 \sum_{{\bf k}{\bf p}}
\delta\left(\hbar \omega_{\bf q}-E_{{\bf p}\downarrow}+E_{{\bf k}\uparrow}\right)
\theta(E_F-E_{{\bf k}\uparrow})\theta(E_{{\bf p}\downarrow}-E_F)g({\bf k})h({\bf p})
\nonumber \\
=\left<g({\bf k})\right>_{\uparrow}\left<h({\bf k})\right>_{\downarrow}\rho_{\uparrow} \rho_{\downarrow}\hbar \omega_{\bf q}
\label{eq:48}
\end{eqnarray}
where $\langle g({\bf k})\rangle_{\sigma}=N^{-1}\sum_{\bf k}g({\bf k})\delta(E_F-E_{{\bf k}\sigma})$ is an 
average over the Fermi surface, as used previously in section 3.1. After some algebra we find

\begin{equation}
 \alpha'_1=2B\alpha/\Delta l_{ex}
\label{eq:49}
\end{equation}

\begin{eqnarray}
 \alpha'_2&=&{\pi c \over \left<n_{\uparrow}-n_{\downarrow}\right>l_{ex}^2\Delta^2}
\Bigg\{\rho_{\uparrow} \rho_{\downarrow}\left(u^2+5v^2\overline{\cos^2\theta}\right)
\sum_{\sigma}\left<\left({\partial \epsilon_{\bf k} \over \partial k_z}\right)^2\right>_{\sigma}
\nonumber \\
&-&2\rho_{\uparrow} \rho_{\downarrow} \Delta v^2 \overline{\cos^2\theta}
\sum_{\sigma}\sigma\left<{\partial^2 \epsilon_{\bf k} \over \partial k^2_z}\right>_{\sigma}
\nonumber \\
&-&v^2\overline{\sin^2\theta}\left[\Delta\sum_{\sigma}\sigma\rho^2_{\sigma}\left<{\partial^2 \epsilon_{\bf k} \over \partial k^2_z}\right>_{\sigma}
-3\sum_{\sigma}\rho^2_{\sigma}\left<\left({\partial \epsilon_{\bf k} \over \partial k_z}\right)^2\right>_{\sigma}\right]
\Bigg\} 
\nonumber \\
&+&{2D \alpha \over \Delta l^2_{ex}}.
\label{eq:50}
\end{eqnarray}
We note that, unlike $\alpha$ and $\alpha'_1$, the coefficient $\alpha'_2$ is non-zero even when the spin-dependent
 part of the impurity potential, $v$, is zero. In this case the damping of a spin wave of frequency $\omega$ 
and small wave-vector $q$ is proportional to $\rho_{\uparrow}\rho_{\downarrow}u^2\omega q^2$. In zero external
field $\omega \sim q^2$ so that the damping is of order $q^4$. This damping due to spin-independent
potential scattering by impurities was analysed in detail by Yamada and Shimizu \cite{ref:15}. One of the
Fermi surface averages in eq.(\ref{eq:50}) is easily evaluated using eqs.(\ref{eq:12b}) and (\ref{eq:16}). Thus
\begin{equation}
\left<{\partial^2 \epsilon_{\bf k} \over \partial k^2_z}\right>_{\sigma}=
-{a^2_0 \over 3}\left<\epsilon_{\bf k}\right>_{\sigma}=-{a^2_0 \over 3}\left(E_f-U\left<n_{-\sigma}\right>+\sigma\mu_B B_{ext}\right).
\label{eq:51}
\end{equation}
In the spirit of the LLG equation we should take $B_{ext}=0$ in evaluating the coefficients $\alpha'_2$.

We now turn to the evaluation of the non-adiabatic spin-transfer torque coefficients $f,f_1$ and $g_1$. These arise
from the second and third terms in square brackets in eq.(\ref{eq:39}), and in a similar expression for $T_2$.
The summations involved in these terms differ from those in eq.(\ref{eq:48}) since one $\theta$-function is
 replaced by a $\delta$-function. This leads to the omission of the frequency factor $\hbar\omega_{\bf q}$.
The Fermi surface shifts $\delta_{\sigma}$ are elininated in favour of currents $J_{\sigma}$ by using eq.(\ref{eq:23}). 

By comparing the coefficient of $q$ in the expansion of eq.(\ref{eq:35}) with that in eq.(\ref{eq:10}) we find 
the coefficient of the Zhang-Li torque in the form 
\begin{equation}
 f={\pi c v^2 \over \mu_0m^2_s\Delta l_{ex}}{\hbar \over e}
\left[2\overline{\cos^2\theta}\left(\rho_{\uparrow} J_{\downarrow}-\rho_{\downarrow} J_{\uparrow}\right)
+\overline{\sin^2\theta}\left(\rho_{\downarrow} J_{\downarrow}-\rho_{\uparrow} J_{\uparrow}\right)\right].
\label{eq:52}
\end{equation}
This is in agreement with Khono {\it et al} \cite{ref:8} and Duine {\it et al} \cite{ref:14}. In the
``isotropic'' impurity case, with $\overline{\cos ^2 \theta}=1/3$, $\overline{\sin ^2 \theta}=2/3$, it
follows from eqs.(\ref{eq:52}), (\ref{eq:47}) and (\ref{eq:2}) that
\begin{equation}
 \beta={f \over a}=\alpha \;\;{2 \over U \left(\rho_{\uparrow}+\rho_{\downarrow}\right)}.
\label{eq:53}
\end{equation}
In the limit of a very weak itinerant forromagnet $\rho_{\sigma}\rightarrow\rho$, the paramagnetic density of states, and $U\rho \rightarrow1$ by the Stoner criterion. Thus in this limit $\beta=\alpha$. 
Tserkovnyak {\it et al} \cite{ref:10} reached a similar conclusion. For a parabolic band it is 
straightforward to show from Stoner theory that $\beta/\alpha>1$ and may be as large as 1.5.
 
As discussed in section 2 the coefficient 
$f_1$ is associated with spin-conserving processes, and hence involves the spin independent potential $u$.
The coefficient $g_1$ is associated with spin non-conserving processes and involves $v$. By comparing the
coefficient of $q^3$ in the expansion of eq.(\ref{eq:35}) with that in eq.(\ref{eq:10}) we deduce that
\begin{equation}
 f_1={ \pi c \over 2\mu_0 m^2_s l^3_{ex}}u^2\left(K_1+2L_1+M_1\right)
\label{eq:54}
\end{equation}
and
\begin{eqnarray}
 g_1={1 \over l^2_{ex}} \left({3D \over \Delta} -{a^2_0 \over 6}\right)f
\nonumber \\
+{\pi c v^2 \over2  \mu_0 m^2_s l^3_{ex}}\left[\overline{\cos^2\theta}\left(5K_1+6L_1-M_1\right)
+\overline{\sin^2\theta}\left(3K_2+4L_2\right)\right].
\label{eq:55}
\end{eqnarray}
Here
\begin{eqnarray}
 K_1&=&K_{\downarrow}\rho_{\uparrow}+K_{\uparrow}\rho_{\downarrow}\;\;,\;\;K_2=K_{\downarrow}\rho_{\downarrow}+K_{\uparrow}\rho_{\uparrow}
\nonumber \\
L_1&=&L_{\downarrow}\rho_{\uparrow}-L_{\uparrow}\rho_{\downarrow}\;\;,\;\;L_2=L_{\downarrow}\rho_{\downarrow}-L_{\uparrow}\rho_{\uparrow}
\nonumber \\
M_1&=&{\hbar \over e \Delta^3}\sum_{\sigma}\left[2\sigma\left<\left({\partial \epsilon_{\bf k} \over \partial k_z}\right)^2\right>_{-\sigma}+\Delta \left<{\partial^2 \epsilon_{\bf k} \over \partial k^2_z}\right>_{-\sigma}\right]J_{\sigma}\rho_{-\sigma}.
\label{eq:56}
\end{eqnarray}
This complete the derivation of expressions for all the LLG coefficients of eq.(\ref{eq:1}) within the 
present impurity model 

\section{The extended LLG equation applied to current-driven domain wall motion}
In a previous paper \cite{ref:9} we introduced the $a_1$ and $f_1$ terms of the extended  LLG equation (cf.
 eqns.(\ref{eq:1}), (\ref{eq:2a}) and (\ref{eq:3})) in order to describe numerically-calculated spin-transfer torques acting on a domain wall when it is traversed by an electric current. In that work the origin of the small $f_1$ term for a pure ferromagnetic metal was specific to the domain wall problem; it was shown to be associated with those electronic states at the bulk Fermi surface which decay exponentially as they enter the wall. The analytic derivation of $f_1$ in section 3 (see eqn.(\ref{eq:54})) is based on impurity scattering in the bulk ferromagnet and applies generally to any slowly-varying magnetization configuration. For a ferromagnetic alloy such as permalloy both mechanisms should contribute in the domain wall situation but the impurity contribution would be expected to dominate.

To describe a domain wall we must add to the right-hand side of eqn.(\ref{eq:1}) anisotropy terms of the form
\begin{equation}
-\left({\bf s} \cdot{\bf e}_y\right){\bf s} \times {\bf e}_y+
b^{-1}\left({\bf s}\cdot{\bf e}_z\right){\bf s}\times{\bf e}_z, 
 \label{eq:57}
\end{equation}
where ${{\bf e}}_y$  is a unit vector perpendicular to the plane of the wire. The first term corresponds to easy-plane shape anisotropy for a wire whose width is large compared with its thickness  and the second term arises from a uniaxial field $H_{u}$ along the wire, so that $b=m_s/H_u$. The solution of eqn.(\ref{eq:1}), with the additional terms (\ref{eq:57}), for a stationary N\'{e}el wall in the plane of the wire, with zero external field and zero current, is
\begin{equation}
{{\bf s}} = (sech(z/b^{1/2}),\;\;\;  0,\;\;\;   -tanh(z/b^{1/2})).
\label{eq:58}
\end{equation}
As pointed out in ref. \cite{ref:9} there is no solution of the LLG equation of the form 
${{\bf s}} ={\bf F}(z-v_W t)$, corresponding to a uniformly moving domain wall, when the $f_1$ term is included. It is likely that the wall
 velocity oscillates about an average value, as predicted by Tatara and Kohno \cite{ref:16,ref:17} 
for purely adiabatic torque above the critical current density for domain wall motion. However, 
we may estimate the average velocity $v_W$ using the method of ref. \cite{ref:9}. The procedure 
is to substitute the approximate form  ${{\bf s}} ={\bf F}(z-v_W t)$  in the extended LLG equation (\ref{eq:1}), 
with the terms (\ref{eq:57}) added, take the scalar product with ${\bf F}\times {\bf F}'$  and integrate with 
respect to $z$ over the range $(-\infty, \infty)$. The boundary conditions appropriate to the wall
 are ${{\bf s}}\rightarrow \mp {{\bf e}}_z$  as $z\rightarrow \pm \infty$. Hence for $b_{ext} =0$ we find the
 dimensionless wall velocity to be
\begin{equation}
v_{W}={f\int_{-\infty}^{\infty}\left({\bf F}\times {\bf F}'\right)^2dz+
f_1\int_{-\infty}^{\infty}\left({\bf F}\times {\bf F}''\right)^2dz+
g_1\int_{-\infty}^{\infty}\left({\bf F}''\right)^2dz \over
\alpha \int_{-\infty}^{\infty}\left({\bf F}\times {\bf F}'\right)^2dz+
\alpha'_2\int_{-\infty}^{\infty}\left({\bf F}''\right)^2dz}.
\label{eq:59}
\end{equation}
To estimate the integrals we take ${\bf F}(z)$ to have the form of the stationary wall ${{\bf s}}(z)$ 
(eqn.(\ref{eq:58})) and, with the physical dimensions of velocity restored, the wall velocity is given approximately by
\begin{equation}
v_W=v_0{\beta \over \alpha}{1+f_1(3fb)^{-1} \over 1+\alpha'_2(\alpha b)^{-1}}
\label{eq:60}
\end{equation}
where $v_0 = \mu_B P J/(m_s e)$.  We have neglected $g_1$ here because, like $f$ and $\alpha$, it depends on spin-orbit coupling but is a factor $(a_0/l_{ex})^2$ smaller than $f$ (cf. eqns.(\ref{eq:52}) and (\ref{eq:55})). $f_1$ and $\alpha'_2$ are important because they do not depend on spin-orbit coupling.

It is interesting to compare $v_W$ with the wall velocity observed in permalloy nanowires by Hayashi {\it et al} \cite{ref:18}. We first note that $v_0$ is the velocity which one obtains very simply from spin angular momentum conservation if the current-driven wall moves uniformly without any distortion such as tilting out of the easy plane and contraction \cite{ref:19}. This is never the case, even if $f_1=0$, $\alpha'_2 = 0$, unless 
$\beta = \alpha$. For a permalloy nanowire, with $\mu_0 m_s= 1$ T,  
$v_0 =110P$ m/s for $J=1.5 \cdot 10^8$A/cm$^2$. Thus, from the standard theory with $f_1=0$, $\alpha'_2 = 0$,
 $v_W = 110P\beta/\alpha$ m/s for this current density. In fact Hayashi {\it et al} \cite{ref:18} 
measure a velocity of 110 m/s which implies $\beta>\alpha$ since the spin polarization $P$ is 
certainly less than 1. They suggest that $\beta$ cannot exceed $\alpha$ and that some additional
 mechanism other than spin-transfer torque is operating. However in the discussion following 
eqn.(\ref{eq:53}) we pointed out that in the model calculations it is possible to have $\beta>\alpha$. 
Even if this is not the case in permalloy we can still have $v_W > v_0$ if the last factor in 
eqn.(\ref{eq:60}) is greater than 1 when $f_1$ and $\alpha'_2$ are non-zero. We can estimate terms in 
this factor using the observation from ref. \cite{ref:18}, that $l_W = l_{ex} b^{1/2}  = 23$ nm, 
where $l_W$ is the width of the wall. From eqns.(\ref{eq:52}) and (\ref{eq:54}) we find 
$f_1/(fb) \sim (u/v)^2 (k_F l_W)^{-2}$, where $k_F$ is a Fermi wave-vector. In permalloy we have
 Fe impurities in Ni so that in the impurity potential $u + {\bf v}\cdot{\bf \sigma}$ we estimate $u\sim$ 1 eV
 and $v\sim$ 0.005 eV. The value for $v$ is estimated by noting that the potential ${\bf v}\cdot{\bf \sigma}$ is
 intended to model spin-orbit coupling of the form $\xi {\bf L}\cdot{\bf \sigma}$ with $\xi \lesssim 0.1$ eV and
 $\langle L_z\rangle_{Fe}\sim$ 0.05, $L_z$ being the component of orbital angular momentum in the
 direction of the magnetization \cite{ref:20}. Hence $u/v\sim 200$ and $k_F l_W\sim200$ so
 that $f_1 / (fb)\sim 1$. $ \alpha'_2 / (\alpha b)$ is expected to be of similar magnitude. We conclude
 that the $\alpha'_2$ and $f_1$ terms in the LLG equation (\ref{eq:1}) can be important in domain
 wall motion and should be included in micromagnetic simulations such as OOMMF  \cite{ref:21}.
 For narrower domain walls these terms may be larger than the Gilbert damping $\alpha$ and
 non-adiabatic spin- transfer torque $f$ terms which are routinely included.  Reliable estimates
 of their coefficients are urgently required using realistic multiband models of the ferromagnetic
 metal or alloy.

\section{Conclusions}

The coefficients of all the terms in an extended LLG equation for a current-carrying ferromagnetic wire have been calculated for a simple model. Two of these ($f_1$ and $\alpha'_2$) are of particular interest since they do not rely on spin-orbit coupling and may sometimes dominate the usual damping and non-adiabatic spin-transfer torque terms. One term ($\alpha_1$) which has been introduced by previous authors is shown rigorously to be zero, independent of any particular model. Solutions of the extended LLG equation for domain wall motion have not yet been found but the average velocity of the wall is estimated. It is pointed out that the $f_1$ and $\alpha'_2$ terms are very important for narrow walls and should be included in micrmagnetic simulations such as OOMMF. It is shown that there is no theoretical reason why the wall velocity should not exceed the simplest spin-transfer estimate $v_0$, as is found to be the case in experiments on permalloy by Hayashi {\it et al} \cite{ref:18}

\ack
\noindent
We are grateful to the EPSRC for financial support through the Spin@RT consortium and to other members of this consortium for encouragement and stimulation.

\appendix
\section{}
The simple single-band impurity model used in the main text is useful for obtaining explicit expressions for 
all the coefficients in the LLG equation (\ref{eq:1}). Here we wish to show that some of these results 
are valid for a completely general system. We suppose the ferromagnetic material is described
by the many-body Hamiltonian
\begin{equation}
 H=H_1+H_{int}+H_{ext}
\label{eq:A1}
\end{equation}
where $H_1$ is a one-electron Hamiltonian of the form
\begin{equation}
 H_1=H_k+H_{so}+V.
\label{eq:A2}
\end{equation}
Here $H_k$ is the total electron kinetic energy, $H_{so}$ is the spin-orbit interaction, $V$ is a potential
term, $H_{int}$ is the coulomb interaction between electrons and $H_{ext}$ is due to an external magnetic
field $B_{ext}$ in the $z$ direction.
Thus 
\begin{equation}
 H_{ext}=-2 \mu_B S^z_0 B_{ext}
\label{eq:A3}
\end{equation}
where $S_z^0$ is the $z$ component of total spin. Both $H_{so}$ and $V$ can contain disorder. Since we are
 interested in the energy and lifetime of a long-wavelength spin wave we consider the spin wave pole,
for small $q$, of the dynamical susceptibility.
\begin{equation}
\chi({\bf q},\omega)=\int dt \langle \langle S^-_{\bf q}(t),S^+_{\bf -q}\rangle \rangle e^{-i \omega_- t}
\label{eq:A4} 
\end{equation}
$(\omega_-=\omega-i\epsilon)$ where $S^\pm_{\bf q}=S^x_{\bf q}\pm iS^y_{\bf q}$ are Fourier components
of the total transverse spin density. Here 
\begin{equation}
\langle \langle S^-_{\bf q}(t),S^+_{\bf -q}\rangle \rangle ={i \over \hbar}\langle \left[ S^-_{\bf q}(t),S^+_{\bf -q}\right] \rangle \theta(t).
\label{eq:A5} 
\end{equation}
In general we shall take the average $\langle \rangle$ in a steady state in which a charge current  density
$J$ is flowing in the ${\bf q}$ direction. Following the general method of Edwards and Fisher \cite{ref:22}
we use equations of motion to find that
\begin{equation}
\chi({\bf q},\omega)=-{2 \langle S^z_0\rangle \over \hbar \left( \omega-b_{ext}\right)}+{1 \over \hbar^2 \left(\omega-b_{ext}\right)^2}
\left\{\chi_c({\bf q},\omega)-\langle \left[ C^-_{\bf q},S^+_{\bf -q}\right] \rangle \right\}
\label{eq:A6} 
\end{equation}
where $\hbar b_{ext}=2\mu_{B} B_{ext}$, $C^-_{\bf q}=[S^-_{\bf q},H_1]$ and
\begin{equation}
\chi_c(q,\omega)=\int dt \langle \langle C^-_{\bf q}(t),C^+_{\bf -q}\rangle \rangle e^{-i \omega t}.
\label{eq:A7} 
\end{equation}
For small $q$ and $\omega$, $\chi$ is dominated by the spin wave pole, so that
\begin{equation}
\chi(q,\omega)=-{2 \langle S^z_0\rangle \over \hbar \left( \omega-b_{ext}-\omega_q\right)}
\label{eq:A8} 
\end{equation}
where $b_{ext}+\omega_{q}$ is the spin wave frequency, in general complex corresponding to a finite lifetime.
Following ref. \cite{ref:22} we compare (\ref{eq:A6}) and (\ref{eq:A8}) in the limit $\omega_q\ll\omega -b_{ext}$
to obtain the general result
\begin{equation}
\omega_q=-{1 \over 2 \langle S^z_0\rangle \hbar}\left\{\lim_{\omega \rightarrow b_{ext}}\chi_c(q,\omega)-\langle \left[ C^-_{\bf q},S^+_{\bf -q}\right] \rangle \right\}.
\label{eq:A9} 
\end{equation}
Edwards and Fisher \cite{ref:22} were  concerned with $Re\, \omega_q$ whereas Kambersky \cite{ref:23} derived the 
above expression for  $Im\, \omega_q$ for the case $q=0$, and zero current flow. His interest was Gilbert damping
in ferromagnetic resonance. Essentially the same result was obtained earlier in connection with electron spin
resonance, by Mori and Kawasaki \cite{ref:24}, see also Oshikawa and Affleck \cite{ref:25}. Since 
$S^-_{\bf q}$ commutes with the potential  term $V$, even in the presence of disorder, we have
\begin{equation}
 C^-_{\bf q}=\left[S^-_{\bf q},H_1\right]=\left[S^-_{\bf q},H_k\right]+\left[S^-_{\bf q},H_{so}\right].
\label{eq:A10}
\end{equation}
For simplicity we now neglect spin-orbit coupling so that
\begin{equation}
 C^-_{\bf q}=\left[S^-_{\bf q},H_k\right]=\hbar q J^-_{\bf q}
\label{eq:A11}
\end{equation}
where the last equation defines the spin current operator $J^-_{\bf q}$. For a general system, 
with the $n^{th}$ electron at position ${\bf r}_n$ with spin $\bf \sigma_n$  and momentum ${\bf p}_n$,   
\begin{equation}
 S^-_q=\sum_n e^{i{\bf q}\cdot {\bf r}_n}\sigma^-_n \;\;,\;\;H_k=\sum_n {\bf p}^2_n /2m.
\label{eq:A12}
\end{equation}
Hence, from eqns.(\ref{eq:A11}) and (\ref{eq:A12}),
\begin{equation}
 \langle \left[ C^-_{\bf q},S^+_{\bf -q}\right] \rangle=N{\hbar^2q^2 \over 2m}+2\hbar\sum_n\langle\sigma^z_n{\bf v}_n\rangle\cdot {\bf q}
\label{eq:A13}
\end{equation}
where $N$ is the total number of electrons and ${\bf v}_n={\bf p}_n/m$ 
is the electron velocity, so that $e\sum_n\langle\sigma^z_n{\bf v}_n\rangle$
is the total spin current. Hence
from eq.(\ref{eq:A9}), we find   
\begin{equation}
\omega_q={\hbar q^2 \over 2 \langle S^z_0\rangle}\left[{N \over 2m}-\lim_{\omega \rightarrow b_{ext}}\chi_J(0,\omega) \right]+{Bq \over \hbar}
\label{eq:A14} 
\end{equation}
with
\begin{equation}
 B=\hbar \mu_B PJ / e m_s.
\label{eq:A15}
\end{equation}
This expression for $B$  has been obtained by Bazaliy {\it et al} \cite{ref:2} and Fern\'{a}ndez-Rossier {\it et al} \cite{ref:13}
for simple parabolic band, $s-d$ and Hubbard models. The derivation here is completely general for any 
ferromagnet, even in the presence of disorder due to impurities  or defects, as long as spin-orbit coupling is
neglected. Eqs.(\ref{eq:2}) and (\ref{eq:A14}) are both valid for arbitrary $b_{ext}$, so that in eq. (\ref{eq:32})
we must have $\alpha_1=0$.

\end{document}